\documentclass[reprint,amsmath,amssymb,aps,pra]{revtex4-1}

\usepackage{hyperref}
\usepackage{graphicx}
\usepackage{dcolumn}
\usepackage{bm}
\usepackage{colortbl}
\usepackage{url}
\usepackage{afterpage}
\usepackage{adjustbox}
\usepackage{placeins}
\begin{document}

\title{Dichroic Filter Characterizations}

\author{A. Bacon}
 \email{ajbacon@sas.upenn.edu}
\author{B. Harris}
\author{J.R. Klein}

\affiliation{Department of Physics and Astronomy, University of Pennsylvania, 209 South 33rd Street Philadelphia PA 19104, USA}

\begin{abstract}

We present here measurements and characterizations of several dichroic filters, which are being used more commonly in nuclear and particle physics for photon-detection. The measurements were performed on filters immersed in several media: air, water, and LAB-based liquid scintillator. Measurements of transmission and reflection properties were made at various angles of incidence, and the data presented here can be used to develop detailed optical models for detector simulations.  We find a modified Bragg's Law is a good model for the shift in transmission edge as a function of angle of incidence, across the various media.
\end{abstract}

\maketitle

\section{Introduction}\label{sec:intro}

Dichroic filters selectively transmit and reflect specific wavelengths of light while minimizing absorption. These filters are becoming more widely used in nuclear and particle physics applications: the photon-trapping ARAPUCAs that were used in the ProtoDUNE detector's photon detection system~\cite{Segreto_2018} and are currently planned for use in DUNE's Far Detector modules~\cite{BRIZZOLARI2025170004}, for example, or the spectral photon-sorting dichroicons~\cite{Kaptanoglu_2020}, which have been deployed in the EOS demonstrator~\cite{Anderson_2023} and are under consideration for the large-scale hybrid Theia detector~\cite{2020EPJC...80..416A}.

Dichroic filters are characterized by ``cut-on" and ``cut-off" wavelengths:  a {\it longpass} filter ideally transmits wavelengths longer than a particular cut-on wavelength and reflects those shorter; a {\it shortpass} filter transmits wavelengths shorter than a particular cut-off and reflects those longer.  Here, we define the cut-on or cut-off wavelengths as the point where transmission drops to 50\% of its maximum value, at an incidence angle of 45$^\circ$.

In practice, the behavior of real filters can differ significantly from their idealizations, as the response depends steeply on incidence angle and can have transmission `windows' even deep into the reflection region. Immersion in media of different refractive indices from the filters also has an impact. For photon detectors that may use these filters, it is critical to model them in simulation at precisions that may be as tight as 1\%. Thus, our measurements here are intended to provide useful input to such models, such as the \texttt{RAT-PAC} analysis and simulation package~\cite{Seibert2014}, the \texttt{Chroma}~\cite{chroma} GPU-accelerated ray tracer, or the \texttt{Opticks} package~\cite{SimonC_2017} used with GEANT4.

The filters we examine here come from Knight Optical \cite{knight}, Edmund Optics \cite{edmund}, and atomic-layer deposition (ALD) filters from Raytum Photonics \cite{raytum}, and we provide details of these in Table \ref{tab:filters}. We include there the cut-on/-off wavelengths, shapes, dimensions, manufacturers, and media of measurement for each. In addition to the dichroic filters, we have included for comparison three longpass {\it absorbing} (non-dichroic) filters from Knight Optical and Edmund Optics.

In this work, our objective is to characterize the optical properties of these filters under various conditions and provide accurate data for simulation models.

\begin{table*}[t!]
    \caption{Summary of dichroic filters (and three absorbing filters) characterized in this study. The cut-on or cut-off wavelength corresponds to 50\% transmission at 45$^{\circ}$ incidence in air. Knight Optical filters were custom manufactured to the dimensions listed. Not all table entries are shown in the figures throughout the paper; the full dataset, including those plotted, is available via the GitHub links in \ref{sec:filtercharacterization-spectrophotometerresults}.}
    \label{tab:filters}
    \centering
    \begin{adjustbox}{width=\textwidth}
    \begin{tabular}{cccccccc}
         \hline\hline\noalign{\smallskip}
          Cut-On/-Off (nm) & Pass & Shape & Dimensions (mm) & Quantity & Manufacturer & Medium &  Online Datasheet \\
         \noalign{\smallskip}\hline\noalign{\smallskip}
          510 & Short &   Rectangular   & 25.2 $\times$ 35.6 & 1 & Edmund Optics & air, water & \cite{edmund500sp} \\
          492 & Long &   Rectangular   & 25.2 $\times$ 35.6 & 1 & Edmund Optics & air, water, LAB & \cite{edmund500lp} \\
          454 & Long &   Rectangular   & 25.2 $\times$ 35.6 & 1 & Knight Optical & air, water & \cite{knight500lp} \\
          422 & Short & Square & 25 $\times$ 25 $\times$ 1.1 & 1 & Knight Optical & air, water & \cite{knight450sp} \\ 
         422 & Short & Trapezoid & 50.0 $\times$ 50.0 & 1 & Knight Optical & air, water & \cite{knight450sp_EOS} \\
         455 & Long & Circular & $\boldsymbol{\oslash}$ 50 & 1 & Knight Optical & air & \cite{knight455lp} \\ 
         475 & Long & Circular & $\boldsymbol{\oslash}$ 50 & 1 & Knight Optical & air & \cite{knight475lp} \\ 
         504 & Long & Circular & $\boldsymbol{\oslash}$ 50 & 1 & Edmund Optics & air, water & \cite{edmundlp} \\
         368 & Short & Circular & $\boldsymbol{\oslash}$ 101.6 & 1 & Raytum Photonics & air & \cite{raytum} \\
         390 & Short & Rectangular & 77 $\times$ 100 & 1 & Raytum Photonics & air & \cite{raytum} \\
         408 & Short & Rectangular & 77 $\times$ 100 & 1 & Raytum Photonics & air & \cite{raytum} \\
         \noalign{\smallskip}\hline\hline
    \end{tabular}
    \end{adjustbox}
\end{table*}

\section{Dichroic Filter Characterization}\label{sec:filter-characterization}

Dichroic filter manufacturers typically provide data for limited incidence angles [0$^{\circ}$ and/or 45$^{\circ}$], often measured primarily in air. We have used two setups for our more extensive measurements, a benchtop PMT/LED setup, and a LAMBDA 850+ UV/Vis spectrophotomer~\cite{spectrophotometer}. Both methods of characterization provide information on transmission; for reflection, we rely exclusively on the PMT setup, although these are broader-band measurements than the spectrophotometer transmission curves, given the broad LED spectra. To study the performance of the filters in different media, we placed the filters into a cuvette filled with water or scintillator and performed spectrophotometer transmission measurements at various angles of incidence (Section \ref{sec:filtercharacterization-spectrophotometerresults}).

\subsection{LED Setup}\label{sec:filtercharacterization-setup}

\begin{figure}[b!]
    \centering
    \includegraphics[trim=0.2cm 0.2cm 0.2cm 0.2cm, clip=true, width=0.45\textwidth]{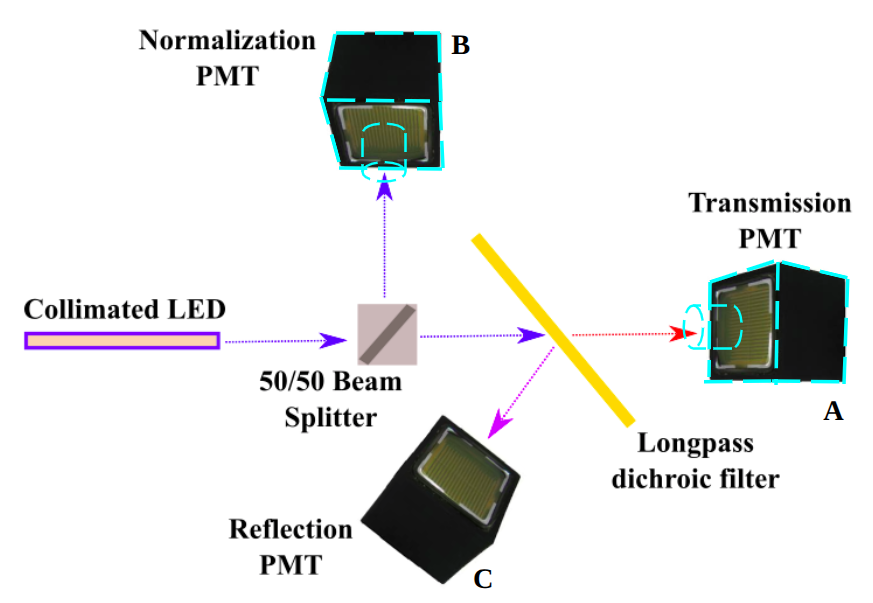}
    \caption{The first dichroic filter characterization test setup. The R7600-U200 PMTs, now denoted ``normalization," ``transmission," and ``reflection" PMTs, collimated LED, and 50/50 beamsplitter are shown. The dichroic filter is held on a rotating stage, not depicted. Not shown is the breadboard holding each structure. The blue dashed lines on the ``transmission" and ``normalization" PMTs represent the collimators used to reduce light leakage and assist with beam alignment. This figure is credited to Kaptanoglu et al. \cite{Kaptanoglu_2020}}
    \label{fig:filter-characterization-setup}
\end{figure}

A diagram of our LED/PMT setup is shown in Figure \ref{fig:filter-characterization-setup}. The setup uses a collimated LED directed towards the center of a 50/50 beamsplitter \cite{beamsplitter}. One ray from the beamsplitter is monitored by an R7600-U200 PMT \cite{r7600pmts}, here called the ``normalization" PMT, to account for variations in LED intensity across data sets. The other ray from the beamsplitter goes into the examined dichroic filter, with transmitted light and reflected light detected by two distinct R7600-U200 PMTs, the ``transmission" and ``reflection" PMTs, respectively. Collimators are added to the ``transmission" and ``normalization" PMTs to minimize light leakage and aid in aligning the incident beam. The maximum uncertainty on the angle of the incident light at the ``transmission" PMT is less than 3 degrees. The three PMTs are operated at -800 V.

To ensure stability from run-to-run, the ``transmission" and ``normalization" PMTs and the beamsplitter were mounted on a 200~mm x 200~mm metal breadboard \cite{breadboard}. 

The dichroic filter is also attached to the breadboard and mounted on a rotating stage \cite{rotation}, allowing precision (1$^{\circ}$) selection of the incidence angles. Our measurements were made in increments of 5$^{\circ}$ (10$^{\circ}$ for Raytum filters) at incidence angles ranging from $\sim$ 0 - 75$^{\circ}$, where 0$^{\circ}$ is normal to the filter surface. Reflection measurements were not made at incidence angles less than 10$^{\circ}$ due to shadowing by parts of the setup of the reflected light. 

The LEDs were purchased from Thorlabs~\cite{thorlabs_leds} and had central emission wavelengths of 385, 405, 450, and 505~nm. The spectral full width at half maximum (FWHM) of the LEDs ranged from 12 to 30~nm; we did not use any additional filters to narrow the wavelength range of the collimated beam for the characterization tests.

We pulsed the LEDs with 40~ns- (50~ns- for Raytum) wide 3.5~V pulses at 1 kHz using a KeySight/Agilent 33500B Waveform Generator~\cite{agilent}. The output of the waveform generator is split in two-ways; one side of the generator is used to trigger the oscilloscope acquisition and the other side goes directly into the LED via an optical fibre. At these settings, the LED created roughly 100 photoelectrons (PEs) per triggered event at the ``normalization" PMT. Typically, either the ``transmission" or ``reflection" PMT detected a similar number of PE as the ``normalization" PMT, depending on whether the wavelength was above or below the cut-on or cut-off for the filter. 

\subsection{DAQ and Data Analysis}\label{sec:data-analysis-1}

Our data acquisition (DAQ) system is a Teledyne LeCroy WaveSurfer 4104HD 1 GHz/12 bit oscilloscope \cite{lecroy}, which digitizes the analog signals coming from each of the three PMTs, as well as the trigger pulse from the KeySight/Agilent waveform generator. Data from the PMTs is sampled every 400~ps/pt in 200~ns long waveforms. 

The \texttt{LeCrunch} software \cite{LaTorre} is used to read out the data from the oscilloscope over an ethernet connection, and the data are formatted into custom \texttt{hdf5} files.

We have a \texttt{C++}-based analysis code, which runs over the \texttt{hdf5} files and calculates the amount of light collected by the ``normalization," ``transmission," and ``reflection" PMTs. Each PMT signal is then integrated and the integrated charge is then converted into the number of photons detected for each triggered event. We then sum this up over the entire data set. 

\subsection{Results}\label{sec:filtercharacterization-results}

To calibrate the relative PMT responses, we take two measurements with no dichroic filter in place for each LED: (1) with the ``transmission" PMT in position A (as shown in Figure~\ref{fig:filter-characterization-setup}), and (2) with the ``reflection" PMT moved from position C to position A. The ``normalization" PMT remains fixed in position B and the beamsplitter is left in place. We then compute the ratio of the integrated waveforms between the ``normalization" PMT and the ``transmission"/``reflection" PMTs to normalize for gain differences and efficiencies between the tubes. These calibration results are then used to normalize the dichroic filter data to an expected intensity corresponding to 100\% transmission or reflection.

We obtain the calculated transmission, $T$, through the dichroic filter by using Equation \ref{eq:transmission},

\begin{equation}\label{eq:transmission}
T = \frac{T_{F}}{T_{NF}} \times \frac{N_{NF}}{N_{F}}
\end{equation}

\begin{figure}[b!]
    \centering
    \includegraphics[trim=0.2cm 0.2cm 0.2cm 0.2cm, clip=true, width=0.45\textwidth]{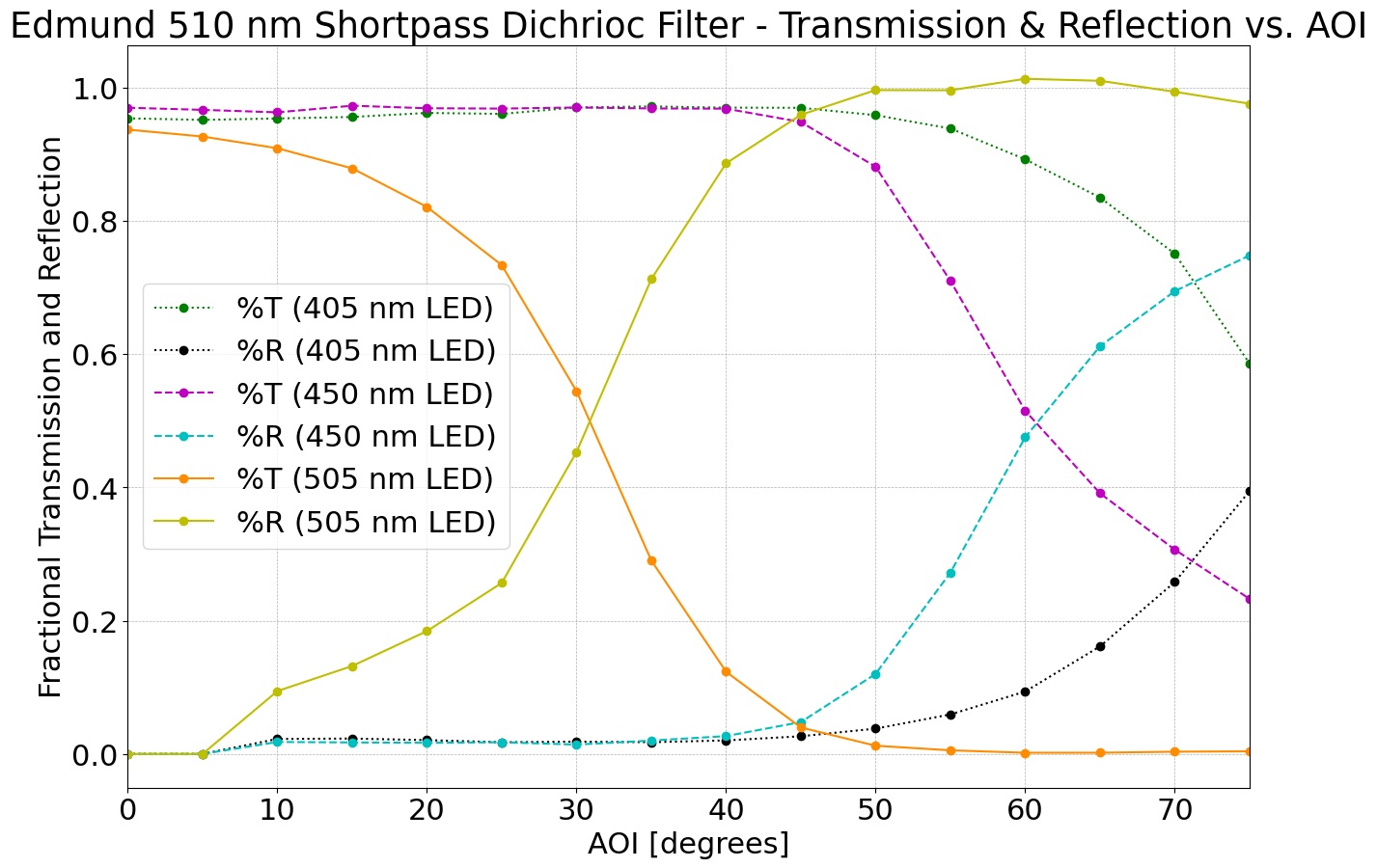}
    \caption{Transmission and reflection data for the Edmund 510~nm shortpass dichroic filter in air, measured for incidence angles between 0$^\circ$ and 75$^\circ$ using 405, 450, and 505~nm LEDs. Transmission and reflection were calculated using Equations \ref{eq:transmission} and \ref{eq:reflection}. The uncertainties are on the scale of the size of the data points, and there is a maximum angular uncertainty of $\sim$3 degrees.}
    \label{fig:Edmund_shortpass_500nm_air_no_cuvette_paper}
\end{figure}

where $T_{F}$ is the total light detected at the ``transmission" PMT, and $T_{NF}$ is the total we obtain from the calibration run with no dichroic filter. The first term is thus the fractional transmittance of the filter under the assumption that the LED intensity remains constant. To account for variations in LED intensity, we multiply by the ratio of the ``normalization" PMT measurement of the LED intensity without the filter present, $N_{NF}$, to the ``normalization" PMT measurement with the filter, $N_{F}$. The reflectance $R$ is calculated the same way but with the data for the ``reflection" PMT instead of the ``transmission" PMT:

\begin{equation}\label{eq:reflection}
R = \frac{R_{F}}{R_{NF}} \times \frac{N_{NF}}{N_{F}}.
\end{equation}

The transmission and reflection results for the 510~nm shortpass dichroic filter from Edmund Optics (in air) are shown in Figure \ref{fig:Edmund_shortpass_500nm_air_no_cuvette_paper}, as a function of incidence angle using three different LEDs: 405 (dotted line), 450 (dashed line), and 505~nm (solid line). The green, magenta, and orange circles represent the transmission as a function of incidence angle, and the black, cyan, and yellow circles depict reflection. As expected, transmission decreases and reflection increases with higher incidence angles--with the transition occurring sooner for wavelengths nearer the shortpass filter's 500~nm cut-off wavelength. Within the precision of our measurement, it also appears that there is little light lost to absorption: $T+R=1$ for each angle. The deviation from unity remains small, with standard deviations of 1.5\%, 1.0\%, and 2.5\% for the 405, 450, and 505~nm LEDs, respectively.

\begin{figure}[b!]
    \centering
    \includegraphics[trim=0.2cm 0.2cm 0.2cm 0.2cm, clip=true, width=0.45\textwidth]{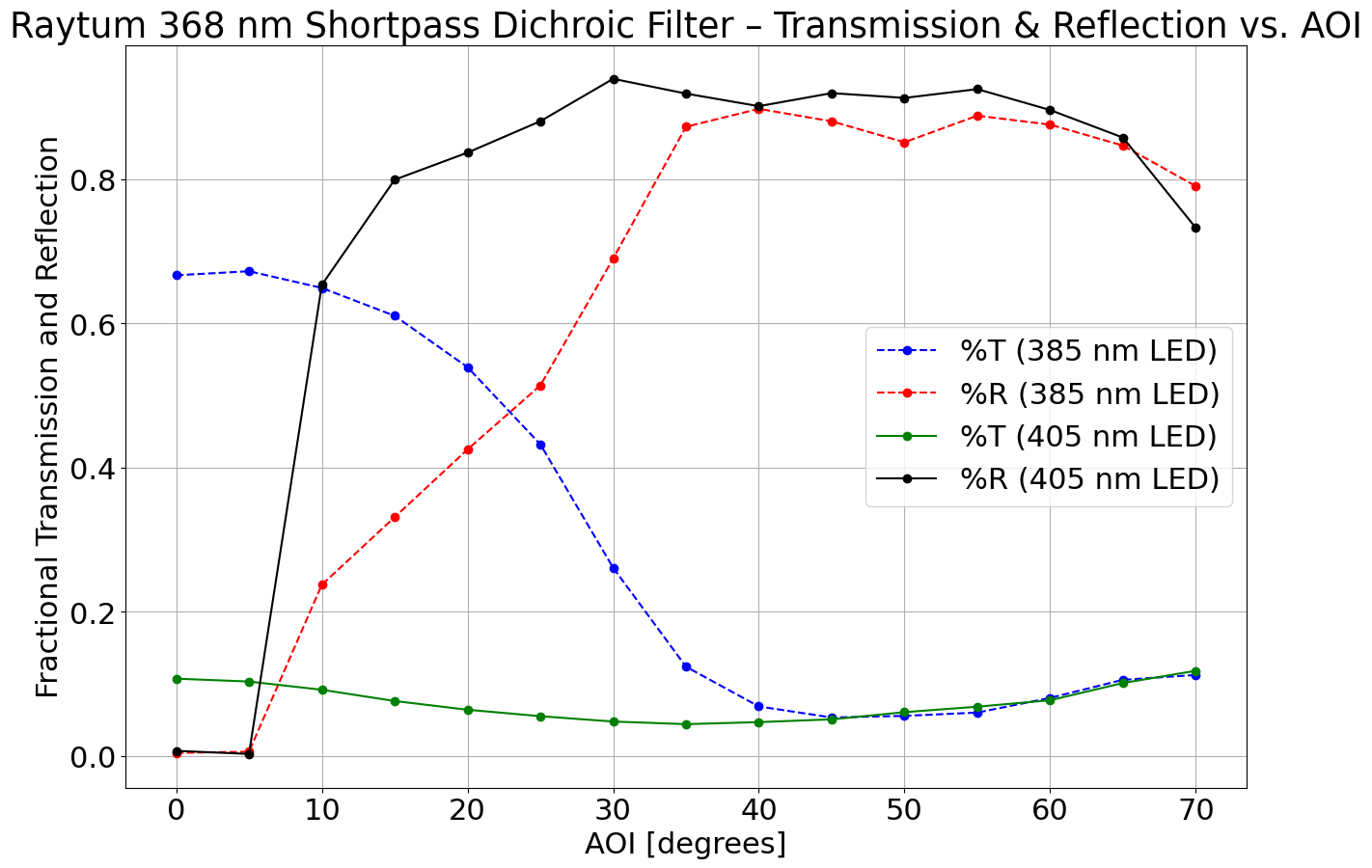}
    \caption{Transmission and reflection data for the Raytum 368~nm shortpass dichroic filter in air, measured for incidence angles between 0$^\circ$ and 70$^\circ$ using 385 and 405~nm LEDs. Transmission and reflection were calculated using Equations \ref{eq:transmission} and \ref{eq:reflection}. The uncertainties are on the scale of the size of the data points, and there is a maximum angular uncertainty of $\sim$3 degrees.}
    \label{fig:Sapphire_4_in_SP_air_T_and_R_385nm_and_405nm_LED_May_15_paper}
\end{figure}

\begin{figure}[b!]
    \centering
    \includegraphics[trim=0.2cm 0.2cm 0.2cm 0.2cm, clip=true, width=0.45\textwidth]{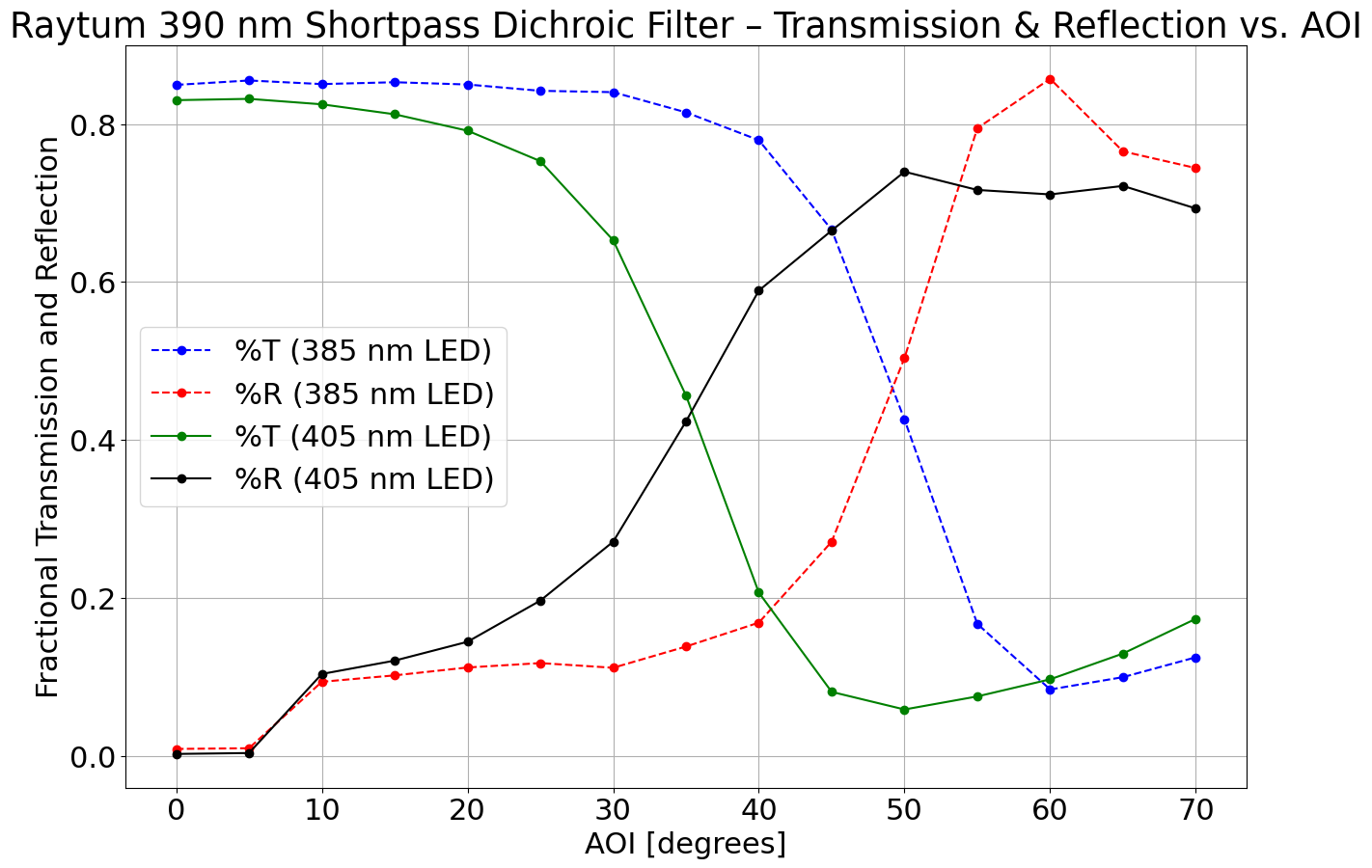}
    \caption{Transmission and reflection data for the Raytum 390~nm shortpass dichroic filter in air, measured for incidence angles between 0$^\circ$ and 70$^\circ$ using 385 and 405~nm LEDs. Transmission and reflection were calculated using Equations \ref{eq:transmission} and \ref{eq:reflection}. The uncertainties are on the scale of the size of the data points, and there is a maximum angular uncertainty of $\sim$3 degrees.}
    \label{fig:Sapphire_B15_SP_air_T_and_R_385nm_and_405nm_LED_may_13_paper}
\end{figure}

\begin{figure}[b!]
    \centering
    \includegraphics[trim=0.2cm 0.2cm 0.2cm 0.2cm, clip=true, width=0.45\textwidth]{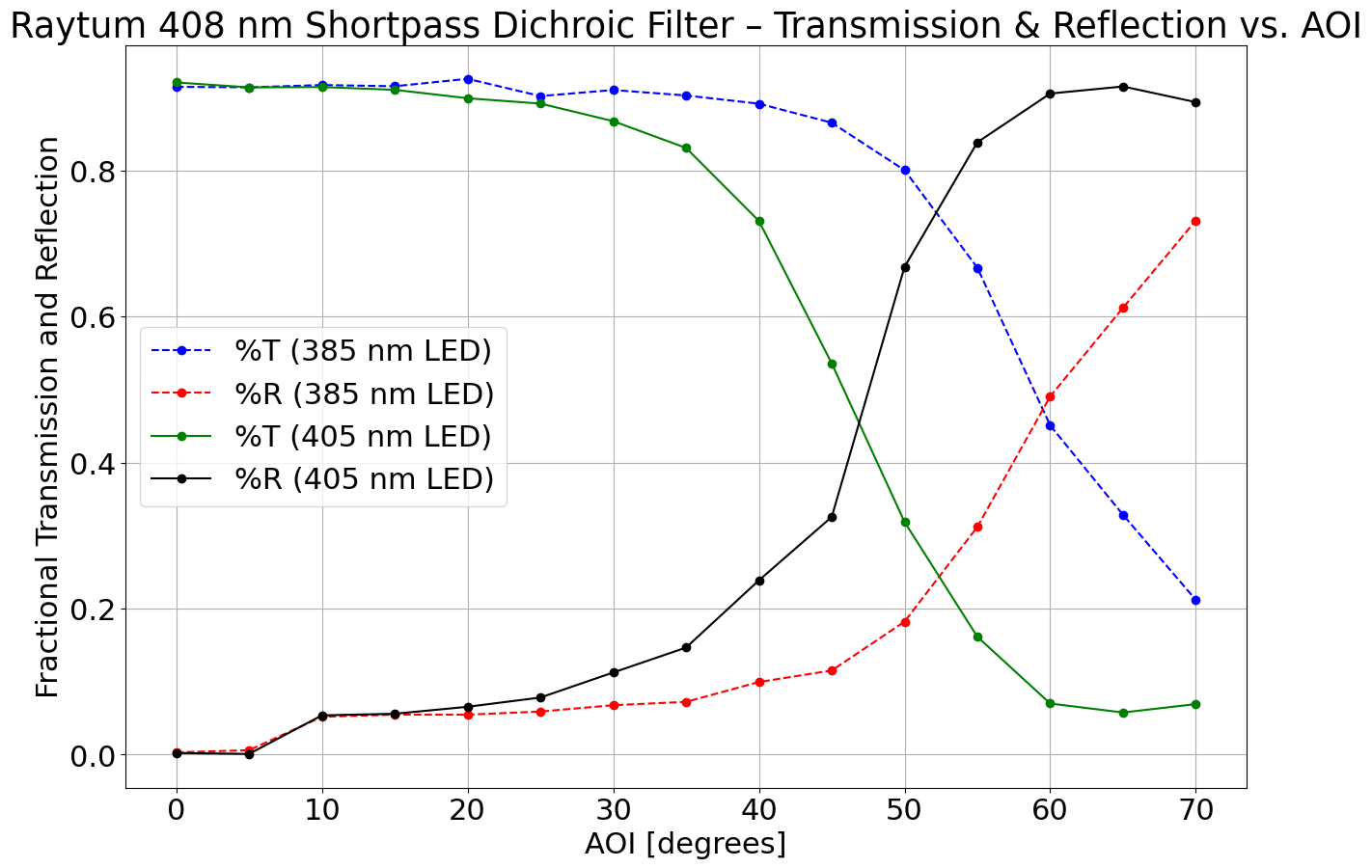}
    \caption{Transmission and reflection data for the Raytum 408~nm shortpass dichroic filter in air, measured for incidence angles between 0$^\circ$ and 70$^\circ$ using 385 and 405~nm LEDs. Transmission and reflection were calculated using Equations \ref{eq:transmission} and \ref{eq:reflection}. The uncertainties are on the scale of the size of the data points, and there is a maximum angular uncertainty of $\sim$3 degrees.}
    \label{fig:B33_B23_SP_air_T_and_R_385nm_and_405nm_LED_May_13_paper}
\end{figure}

For the Raytum shortpass dichroic filters, the transmission and reflection results follow expected trends, with transmission decreasing and reflection increasing at higher incidence angles. The slight rise in transmission at high incidence angles is surprising and appears to be a property of the filter optics. The 368~nm shortpass filter (Figure \ref{fig:Sapphire_4_in_SP_air_T_and_R_385nm_and_405nm_LED_May_15_paper}) shows a more gradual decline in transmission compared to the 390~nm (Figure \ref{fig:Sapphire_B15_SP_air_T_and_R_385nm_and_405nm_LED_may_13_paper}) and the 408~nm (Figure \ref{fig:B33_B23_SP_air_T_and_R_385nm_and_405nm_LED_May_13_paper}) shortpass filters, which exhibit sharper transitions. This behavior suggests a stronger angular dependence for filters with longer cut-off wavelengths. Additionally, the measured values align closely with manufacturer specifications within experimental uncertainties. Within the precision of our measurement, it also appears that there is little light lost to absorption: $T+R=1$ for each angle. For the 368~nm shortpass filter, the summed transmission and reflection initially showed large deviations from unity due to the absence of reflection measurements at 0$^\circ$ and 5$^\circ$, where the reflected beam could not be captured. Excluding these points, the standard deviations were 2.6\% (385~nm LED) and 4.5\% (405~nm LED). The 390~nm shortpass filter exhibited moderately larger deviations, with standard deviations of  3.9\% (385~nm LED) and 6.2\% (405~nm LED). In contrast, the 408~nm shortpass filter showed smaller deviations of 2.3\% (385~nm LED) and 3.4\% (405~nm LED). These discrepancies likely arise from a combination of scattered light, alignment uncertainties, or absorption at higher angles of incidence (AOI).

Only a small fraction of the dichroic filter is illuminated by the light from the beamsplitter, meaning the ``transmission" PMT samples light from a localized region of the filter. To evaluate possible spatial variations in transmission and reflection, we performed a uniformity study using the filters shown in Figures \ref{fig:Sapphire_4_in_SP_air_T_and_R_385nm_and_405nm_LED_May_15_paper}-\ref{fig:B33_B23_SP_air_T_and_R_385nm_and_405nm_LED_May_13_paper}. For Figure \ref{fig:Sapphire_4_in_SP_air_T_and_R_385nm_and_405nm_LED_May_15_paper}, the RMS spread in the 50\% transmission wavelength was 1.7~nm at 0$^{\circ}$ and 1.4~nm at 45$^{\circ}$. For the other two filters (Figures \ref{fig:Sapphire_B15_SP_air_T_and_R_385nm_and_405nm_LED_may_13_paper} and \ref{fig:B33_B23_SP_air_T_and_R_385nm_and_405nm_LED_May_13_paper}), the RMS spread was less than 0.9~nm across all positions and angles tested. These results indicate that spatial variations in cut-off performance are minimal, with all RMS spreads below 2~nm.

\subsection{Spectrophotometer Setup}\label{sec:filtercharacterization-spectrophotometer}

For precision measurements of transmission across different wavelengths and incidence angles, we use a LAMBDA 850+ UV/Vis spectrophotometer \cite{spectrophotometer}. Such spectrophotometers are commonly used for measuring absorbance and transmission in various media, including water, organic scintillators, and water-based liquid scintillators.

The LAMBDA 850+ covers a spectral range of 175–900~nm using a deuterium lamp for the ultraviolet region and a tungsten-halogen lamp for the visible and near-infrared regions. It operates with an intrinsically unpolarized light source. No additional polarizers or depolarizers were introduced into the optical path, ensuring that the measurements reflect the natural unpolarized response of the filters. The spectrophotometer is equipped with a gridless PMT, housed in a separate detection compartment, and provides a high spectral resolution of 0.05~nm.

\begin{figure}[b!]
    \centering
    \includegraphics[trim=0.2cm 0.2cm 0.2cm 0.2cm, clip=true, width=0.45\textwidth]{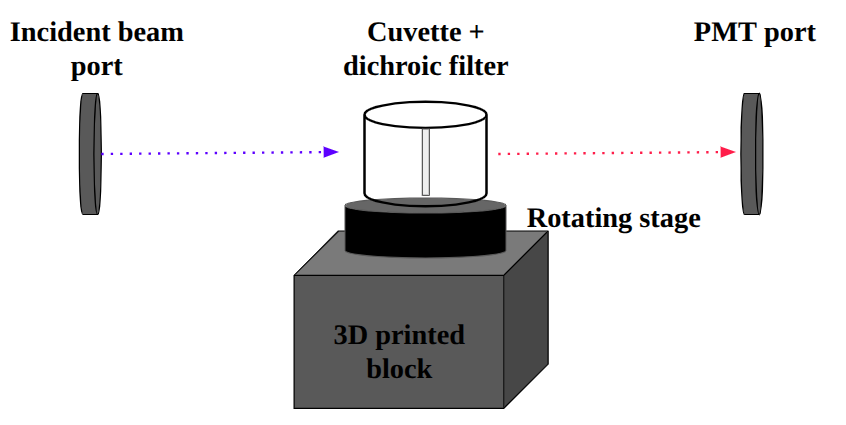}
    \caption{The second dichroic filter characterization test setup. The 3D-printed stand (felt not depicted), rotating stage, cuvette, and dichroic filter are shown. The dichroic filter is held upright by a fixed filter holder, not depicted. Measurements were conducted under three conditions: (1) in air with only the filter in the light path (no cuvette), (2) in air with the filter placed inside an empty cuvette, and (3) in liquid with the filter fully submerged in a cuvette filled with either tap water or LAB. Aside from the cuvette, the rest of the setup remains unchanged between configurations.}
    \label{fig:spm_schermatic}
\end{figure}

For our measurements, we select a wavelength range of 250-900~nm at a resolution of 2~nm, which more than spans the response band of most photon sensors. The dichroic filter is mounted vertically in a filter holder, which is secured to a rotating stage (see Subsection \ref{sec:filtercharacterization-setup}). A 3D-printed stand ensures stability and maintains the appropriate height of the filter. To reduce unwanted reflections or scattering from the plastic material, the stand is covered with black felt. The arrangement is such that incident light passes through the central region of the dichroic filter and is directed towards a port for detection by the spectrophotometer's PMT. A schematic of this setup is depicted in Figure \ref{fig:spm_schermatic}. The cuvette is shown in the schematic for completeness, but the setup is otherwise unchanged when the cuvette is removed.

The rotating stage allows for precise angle adjustments in 1$^{\circ}$ increments. Transmission measurements were taken at incidence angles from $\sim$ 0 - 75$^{\circ}$, with data recorded at 15$^\circ$ intervals. Here, 0$^\circ$ corresponds to normal incidence. Measurements beyond 75$^{\circ}$ were not feasible, as the incident beam began to clip the filter edge, leading to inconsistent transmission readings.

\subsection{Spectrophotometer Results}\label{sec:filtercharacterization-spectrophotometerresults}

After setting up the spectrophotometer-based transmission measurements, we now present the results obtained across various media and angles of incidence.

Each measurement begins with a calibration where the dichroic filter is removed. The measurement is then repeated at the same incidence angle with the filter in place. The calibration allows us to understand the contributions to transmittance from non-dichroic elements of the apparatus.

To investigate the filter's response in different media (e.g., water and linear-alkyl benzene, LAB), we fully submerged the filters inside a 73~mL cylindrical cuvette~\cite{cuvettes} filled with the respective liquid. The cylindrical geometry facilitated measurements at high incidence angles while avoiding complications from edge effects. 

Our ``target out'' calibration runs for each immersion medium included the liquid media in the cuvette. When the dichroic filter was added, it was mounted vertically within the cuvette at the same incidence as the calibration run.

We conducted measurements under three conditions: (1) air without the cuvette, (2) air with the cuvette, and (3) tap water or LAB with the cuvette. Our choice of these media are intended to be representative of what typical photon-based detectors use, with the notable absence of any liquid cryogen, which would be difficult to do within the spectrophotometer.

Our results are summarized in Table \ref{tab:filters} and visualized in Figures \ref{fig:Knight_longpass_500nm_air_tap_water_73mL_cuvette_09_06}–\ref{fig:Bragg_B23_paper}.

As the figures show, there is considerable structure in all of the curves as we get away from 45$^\circ$.

Figure \ref{fig:Knight_longpass_500nm_air_tap_water_73mL_cuvette_09_06} shows the behavior of a 454~nm longpass dichroic filter from Knight Optical measured in both air and tap water (using the 73~mL cylindrical cuvette in both cases). As the AOI increases, there is a noticeable leftward shift in the wavelength at which the filter achieves 50\% transmission. This shift suggests that the filter's performance is influenced by changes in AOI in the presence of air and tap water. It is important to note that shifts are observed between air and tap water cases at all AOIs except at normal incidence. 

Figure \ref{fig:Edmund_shortpass_500nm_air_tap_water_73mL_cuvette_08_31} shows the results for a 510~nm shortpass dichroic filter from Edmund Optics measured in both air and tap water (using the 73~mL cylindrical cuvette in both cases). Similar to the Knight filter, the Edmund filter exhibits a leftward shift in the wavelength of 50\% transmission as AOI increases.  Comparing these results with the first measurement method (Figure \ref{fig:Edmund_shortpass_500nm_air_no_cuvette_paper}) demonstrates consistency in performance trends across different setups. At wavelengths beyond 550~nm, transmission starts to increase again. This is a curious trend we see in nearly every shortpass filter. It appears to be a property of the dichroic filters.

Figure \ref{fig:DF_LP_16_Edmund_500nm_air_LAB_cuvette_05_24} shows a 492~nm longpass dichroic filter from Edmund Optics measured in air and LAB (using the 73~mL cylindrical cuvette in both cases). The shift in wavelength between these media is more pronounced than between air and tap water, due to LAB's higher refractive index. Furthermore, as the AOI increases, the shift from air to LAB gets larger, aligning with the expectations of filters submerged in scintillator or oil. These findings underscore the importance of considering the medium's optical properties when deploying such filters in photon detection systems.

Figure \ref{fig:Absorbing_longpass_500nm_air_tap_water_73mL_cuvette_09_16} displays the results for a 504~nm {\it absorbing} (non-dichroic) longpass filter in air and tap water (using the 73~mL cylindrical cuvette in both cases). We include it here for comparison with the dichroic results, but also because it is used in conjunction with dichroic filters, as in the dichroicon design described in Ref.~\cite{Kaptanoglu_2020}. As expected, the absorbing filter exhibits minimal variation in cut-on wavelength across different AOIs, in contrast to the dichroic filters.

In Figure \ref{fig:Edmund_500nm_longpass_air_LAB_wave_50_percent_error} we use the curves from Figure \ref{fig:DF_LP_16_Edmund_500nm_air_LAB_cuvette_05_24} to show the behavior of the 50\% transmission point as a function of AOI. 
We see that apart from normal incidence, there is a distinct shift between measurements in air and in LAB at each increment of AOI. Beyond 45$^\circ$ in LAB, however, the filter maintained a transmission rate of at least 60$\%$; we do not include those points here. As discussed later, the behavior of this 50\% transmission edge is well-described by a modified Bragg's Law.

Figure \ref{fig:Knight_500nm_longpass_air_h2o_wave_50_percent_error_paper} illustrates the performance of a 454~nm longpass dichroic filter from Knight Optical measured in air and tap water as a function of AOI (data from Figure \ref{fig:Knight_longpass_500nm_air_tap_water_73mL_cuvette_09_06}). The plot again shows the wavelength at which the filter achieves 50$\%$ transmission at each AOI. Similar to the LAB measurements in the previous figure, there is a consistent downward shift in the 50$\%$ transmission wavelength with increasing AOI. This shift is more pronounced in tap water as a result of its higher refractive index compared to that of air. 

Figures \ref{fig:Knight_shortpass_EOS_450nm_air_vs_h2o_wave_50_percent_coated_error_paper} and \ref{fig:Knight_shortpass_EOS_450nm_air_vs_h2o_wave_50_percent_non_coated_error_paper} compare the coated and non-coated sides of a 422~nm shortpass dichroic filter from Knight Optical under varying AOIs in air and tap water (using the 73~mL cylindrical cuvette in both cases). The results show that the coated and non-coated sides exhibit nearly identical transmission behavior, apart from minor instrumental fluctuations. This suggests that for this particular filter, the coating does not significantly alter the AOI dependence in terms of transmission. This filter is used in the dichroicon within EOS, where understanding AOI effects is crucial for optical modeling.

Figure \ref{fig:Knight_500nm_longpass_air_h2o_wave_50_percent_crude_fit_Bragg_bounds_sigma_chi_paper} utilizes data from Figure \ref{fig:Knight_500nm_longpass_air_h2o_wave_50_percent_error_paper}, shown as points with error bars of 2~nm, derived from the spectrophotometer's step size during wavelength scans. The solid lines represent the modified Bragg's Law fit, showing good agreement with the experimental data. The modified Bragg's Law is as follows:

\begin{equation}\label{eq:Bragg}
\lambda = \lambda_{0} \sqrt{1 - \frac{\sin^2(\theta)}{n^2}}
\end{equation}

where $\lambda$ is the observed wavelength at an angle, $\lambda_{0}$ is the wavelength at normal incidence (0$^\circ$), $\theta$ is the AOI, and $n$ is the effective refractive index of the layers. Although we know $\lambda_{0}$ from the data, $\lambda_{0}$ and $n$ are free parameters in our fit. The fit parameters are as follows: for Figure \ref{fig:Knight_500nm_longpass_air_h2o_wave_50_percent_crude_fit_Bragg_bounds_sigma_chi_paper} in air, $\lambda_0 = 498.54$~nm and $n = 1.78$; in water, $\lambda_0 = 494.63$~nm and $n = 1.36$.

Figures \ref{fig:RAYTUM_4_in_SP_air_no_cuvette_coated_2024_07_05_paper}, \ref{fig:RAYTUM_B15_SP_air_no_cuvette_coated_2024_07_05_paper}, and \ref{fig:RAYTUM_B23_SP_air_no_cuvette_coated_2024_07_05_paper} present data for three shortpass dichroic filters from Raytum Photonics, measured in air without a cuvette. As expected, all three filters exhibit a consistent leftward shift in the wavelength at 50$\%$ transmission as AOI increases. The 368~nm shortpass filter shows less angular dependence compared to the 390~nm and 408~nm shortpass filters, with a more gradual shift in the 50$\%$ transmission wavelength across AOI. The 390~nm and 408~nm shortpass filters exhibit steeper shifts, suggesting a stronger AOI dependence, likely due to differences in layer thicknesses, coating designs, or substrate properties. Additionally, the 368~nm and 390~nm shortpass filters show stronger oscillations in their transmission curves compared to the 408~nm shortpass filter, reflecting differences in material properties.

Figures \ref{fig:Bragg_4_in_paper}, \ref{fig:Bragg_B15_paper}, and \ref{fig:Bragg_B23_paper} depict modified Bragg's Law fits for the Raytum filters. Error bars of 2~nm are included. The solid lines represent the fit, highlighting the excellent agreement with the experimental data. The fit parameters are as follows: for Figure \ref{fig:Bragg_4_in_paper} (368~nm filter), $\lambda_0 = 394.93$~nm and $n = 1.97$; for Figure \ref{fig:Bragg_B15_paper} (390~nm filter), $\lambda_0 = 429.30$~nm and $n = 1.70$; and for Figure \ref{fig:Bragg_B23_paper} (408~nm filter), $\lambda_0 = 445.03$~nm and $n = 1.78$. A summary of the filters and fit parameters can be found in Table \ref{tab:Braggs}.

Our results can be found in these \href{https://github.com/hbjamin/dichroic-filter-reflection-tests}{GitHub} \href{https://github.com/amandabacon/dichroic_filter_paper}{repositories}, and can be used in photon detector photon modeling packages, such as \texttt{Chroma}~\cite{chroma,Kaptanoglu_2020}, \texttt{RAT-PAC}~\cite{Seibert2014}, or \texttt{Opticks}~\cite{SimonC_2017}.

\begin{table*}[t!]
    \caption{Summary of dichroic filters fitted by a modified Bragg's Law. The cut-on or cut-off wavelength corresponds to 50\% transmission at 45$^{\circ}$ incidence in air. The manufacturer's quoted cut-on or cut-off wavelength is also included in column 3. The last two columns are the modified Bragg's Law float parameters.}
    \label{tab:Braggs}
    \centering
    \begin{adjustbox}{width=\textwidth}
    \begin{tabular}{ccccc}
         \hline\hline\noalign{\smallskip}
          Cut-On/-Off (nm) & Pass & Manufacture Cut-On/-Off (nm) & $\lambda_0$ (nm) & Effective Index of Refraction (n) \\
         \noalign{\smallskip}\hline\noalign{\smallskip}
          454 & Long & 500 & 498.45 & 1.78 (air) \\
              &      &     & 494.63 & 1.36 (water) \\
          368 & Short & 400 & 394.93 & 1.97 \\
          390 & Short & 430 & 429.30 & 1.70 \\
          408 & Short & 450 & 445.03 & 1.78 \\
         \noalign{\smallskip}\hline\hline
    \end{tabular}
    \end{adjustbox}
\end{table*}

\newpage

\afterpage{
    \begin{figure}[p!]
        \centering
        \includegraphics[trim=0.2cm 0.2cm 0.2cm 0.2cm, clip=true, width=0.60\textwidth]{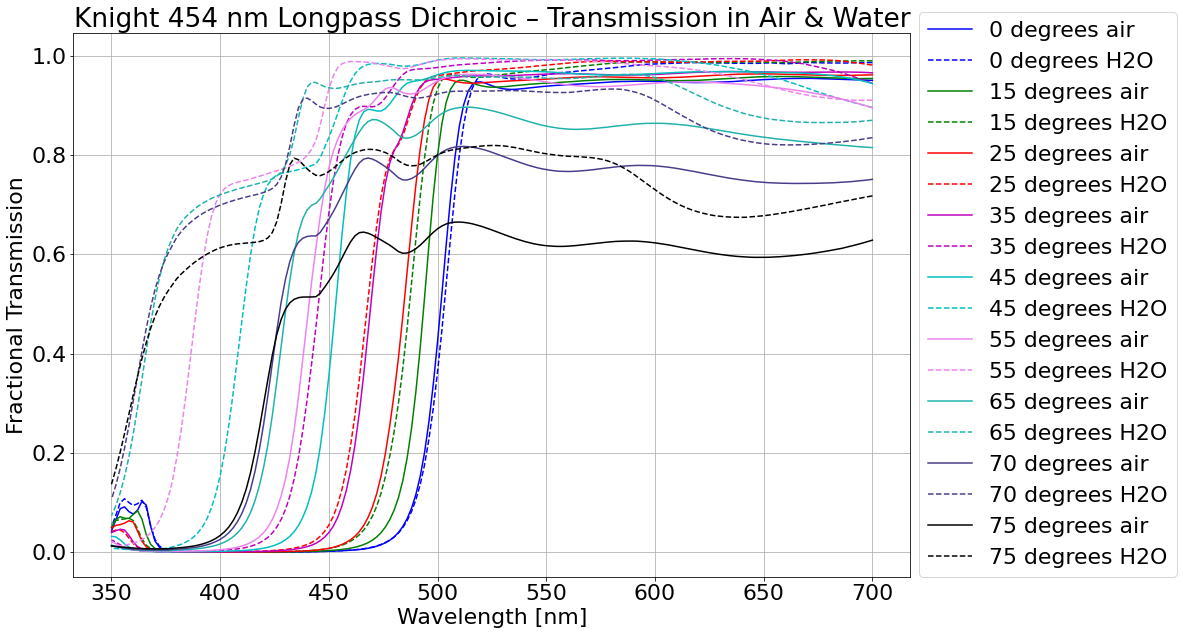}
        \caption{Transmission data (various colors) for the 454~nm longpass filter from Knight Optical, measured in air and tap water for AOIs between 0$^\circ$ and 75$^\circ$ using the spectrophotometer.}
        \label{fig:Knight_longpass_500nm_air_tap_water_73mL_cuvette_09_06}
    \end{figure}

    \vspace{1cm} 

    \begin{figure}[p!]
        \centering
        \includegraphics[trim=0.2cm 0.2cm 0.2cm 0.2cm, clip=true, width=0.60\textwidth]{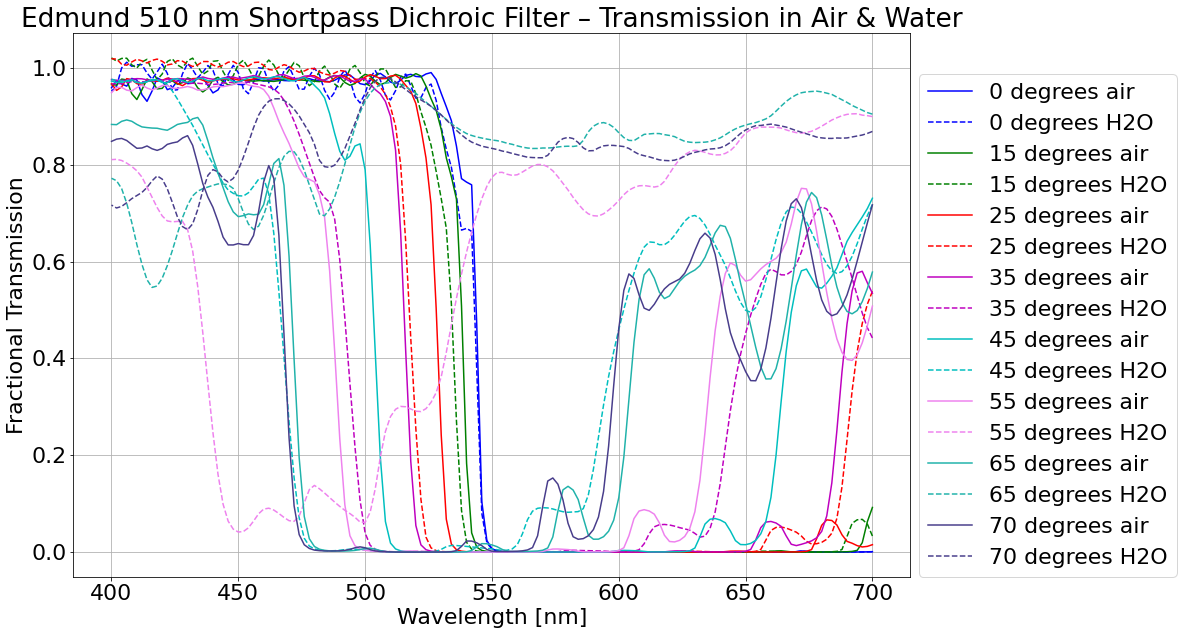}
       \caption{Transmission data (various colors) for the 510~nm shortpass filter from Edmund Optics, measured in air and tap water for AOIs between 0$^\circ$ and 70$^\circ$ using the spectrophotometer.}
        \label{fig:Edmund_shortpass_500nm_air_tap_water_73mL_cuvette_08_31}
    \end{figure}

    \vspace{1cm} 

    \begin{figure}[p!]
        \centering
        \includegraphics[trim=0.2cm 0.2cm 0.2cm 0.2cm, clip=true, width=0.60\textwidth]{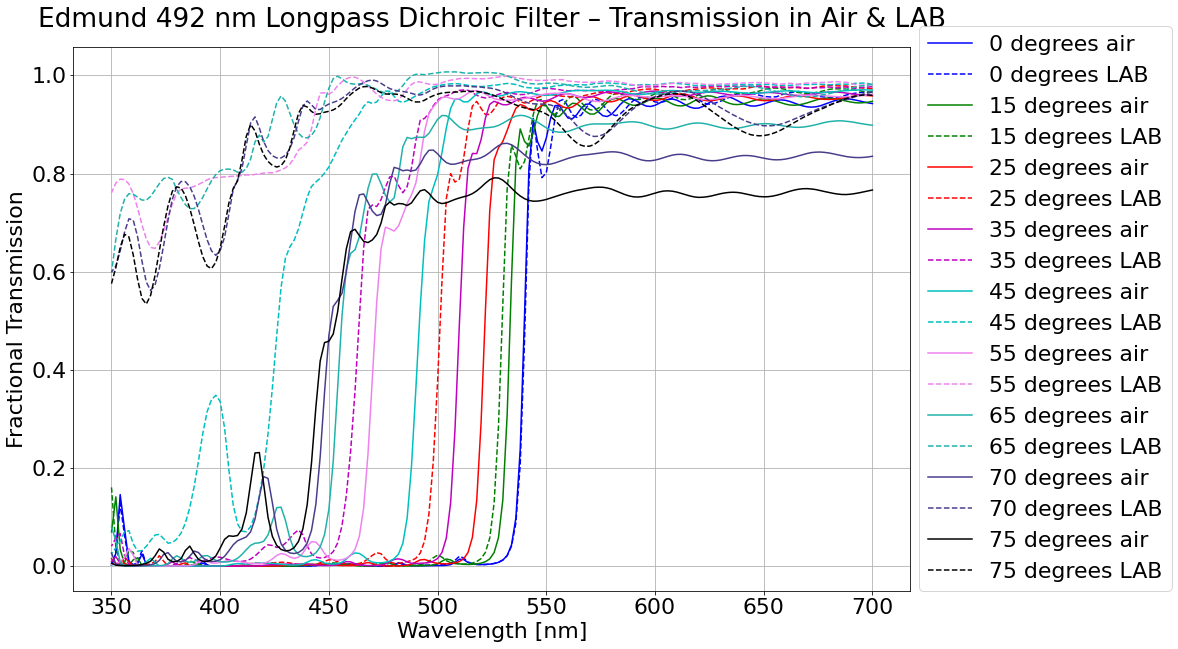}
        \caption{Transmission data (various colors) for the 492~nm longpass filter from Edmund Optics, measured in air and LAB for AOIs between 0$^\circ$ and 75$^\circ$ using the spectrophotometer.}
        \label{fig:DF_LP_16_Edmund_500nm_air_LAB_cuvette_05_24}
    \end{figure}

    \vspace{1cm} 
    \newpage
    \FloatBarrier 
}

\afterpage{
    \begin{figure}[p!]
        \centering
        \includegraphics[trim=0.2cm 0.2cm 0.2cm 0.2cm, clip=true, width=0.60\textwidth]{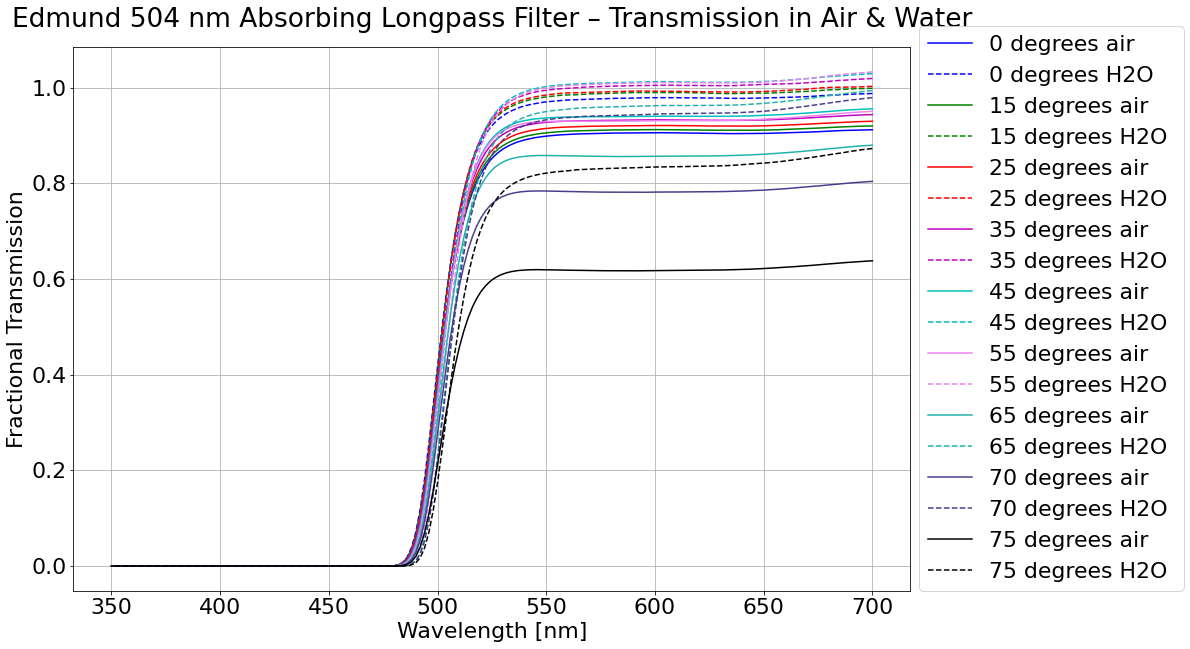}
        \caption{Transmission data (various colors) for the 504~nm absorbing longpass filter from Edmund Optics, measured in air and tap water for AOIs between 0$^\circ$ and 75$^\circ$ using the spectrophotometer.}
        \label{fig:Absorbing_longpass_500nm_air_tap_water_73mL_cuvette_09_16}
    \end{figure}

    \vspace{1cm} 

    \begin{figure}[p!]
        \centering
        \includegraphics[trim=0.2cm 0.2cm 0.2cm 0.2cm, clip=true, width=0.57\textwidth]{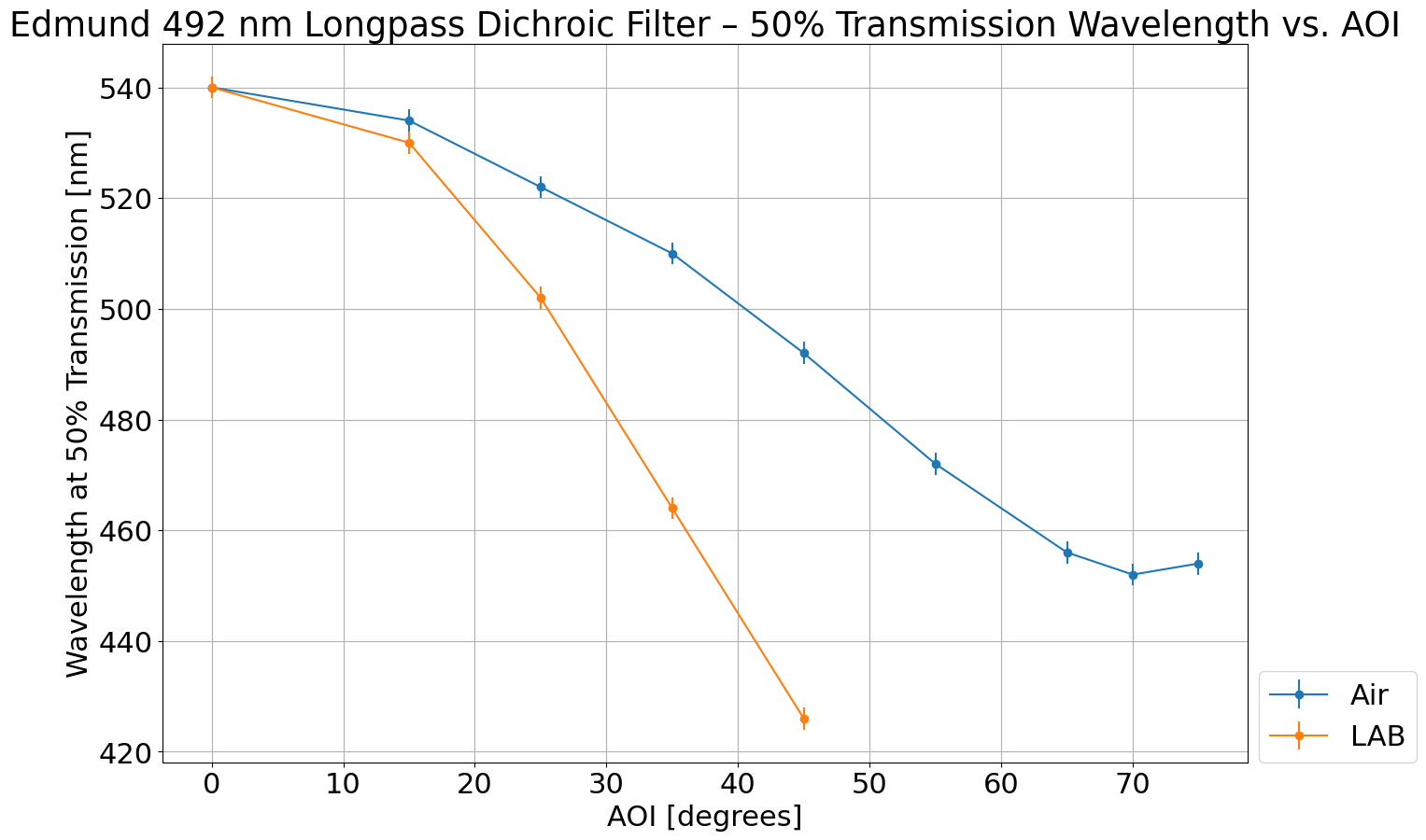}
        \caption{Wavelength at 50$\%$ transmission for the 492~nm longpass filter from Edmund Optics, measured in air and LAB for AOIs between 0$^\circ$ and 75$^\circ$ using the spectrophotometer. Error bars are 2~nm.}
        \label{fig:Edmund_500nm_longpass_air_LAB_wave_50_percent_error}
    \end{figure}

    \vspace{1cm} 
    \newpage
    \FloatBarrier 
}

\afterpage{
    \begin{figure}[p!]
        \centering
        \includegraphics[trim=0.2cm 0.2cm 0.2cm 0.2cm, clip=true, width=0.57\textwidth]{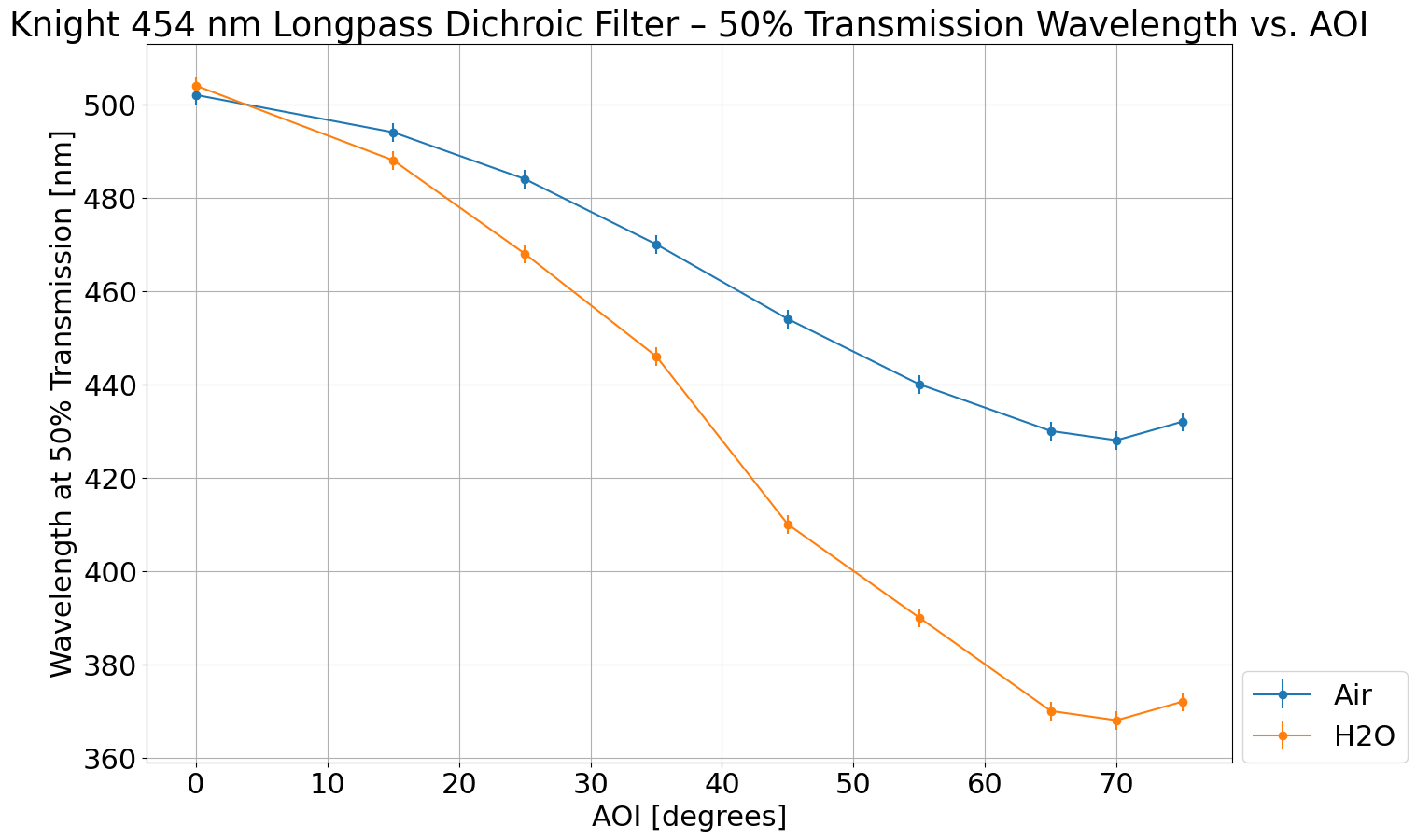}
        \caption{Wavelength at 50$\%$ transmission for the 454~nm longpass filter from Knight Optical, measured in air and tap water for AOIs between 0$^\circ$ and 75$^\circ$ using the spectrophotometer. Error bars are 2~nm.}
        \label{fig:Knight_500nm_longpass_air_h2o_wave_50_percent_error_paper}
    \end{figure}

    \vspace{1cm} 

    \begin{figure}[p!]
        \centering
        \includegraphics[trim=0.2cm 0.2cm 0.2cm 0.2cm, clip=true, width=0.57\textwidth]{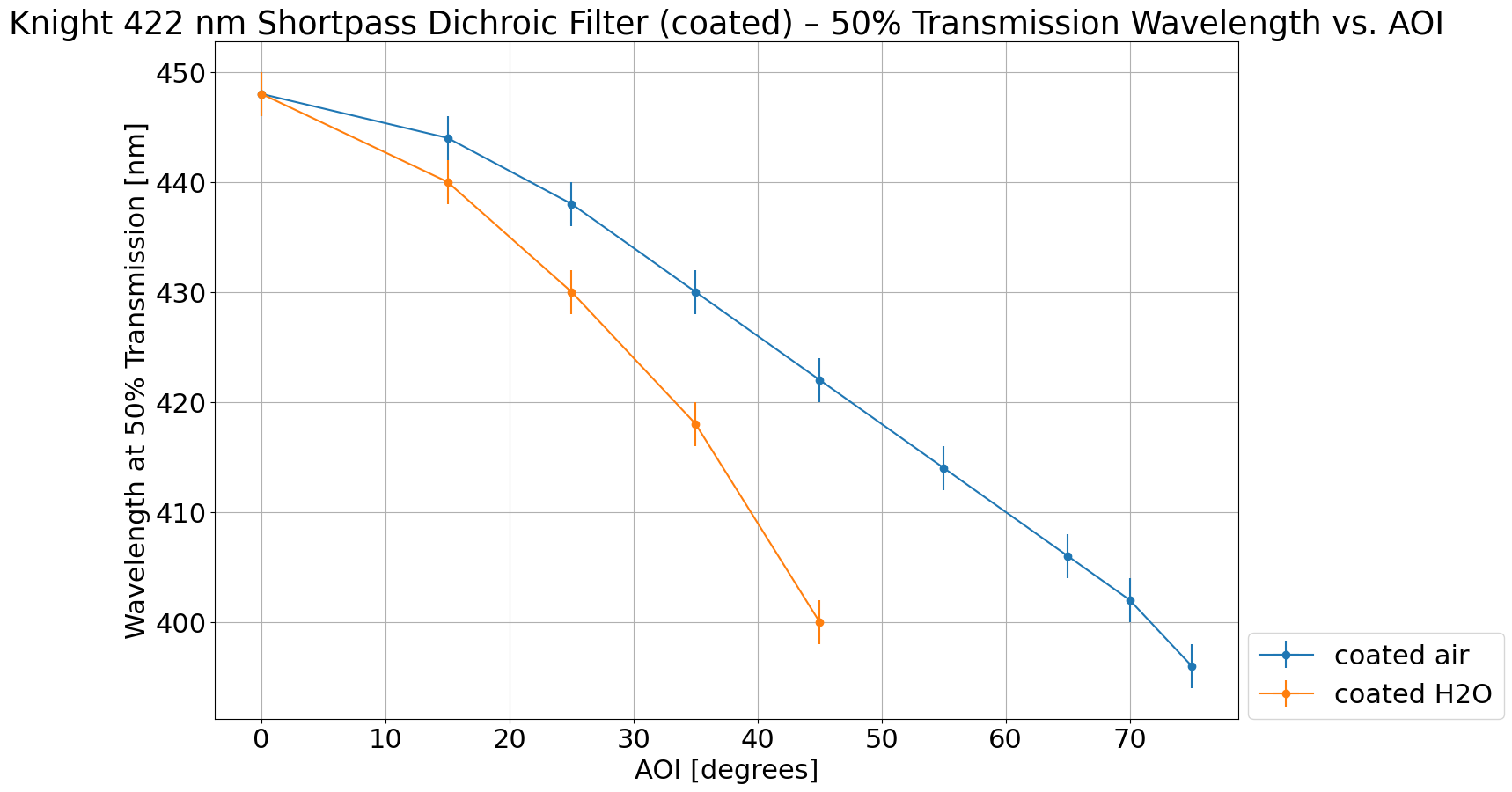}
        \caption{Wavelength at 50$\%$ transmission for the 422~nm shortpass EOS filter (coated side) from Knight Optical, measured in air and tap water for AOIs between 0$^\circ$ and 75$^\circ$ using the spectrophotometer. Error bars are 2~nm.}
        \label{fig:Knight_shortpass_EOS_450nm_air_vs_h2o_wave_50_percent_coated_error_paper}
    \end{figure}

    \vspace{1cm} 
    \newpage
    \FloatBarrier 
}

\afterpage{
    \begin{figure}[p!]
        \centering
        \includegraphics[trim=0.2cm 0.2cm 0.2cm 0.2cm, clip=true, width=0.57\textwidth]{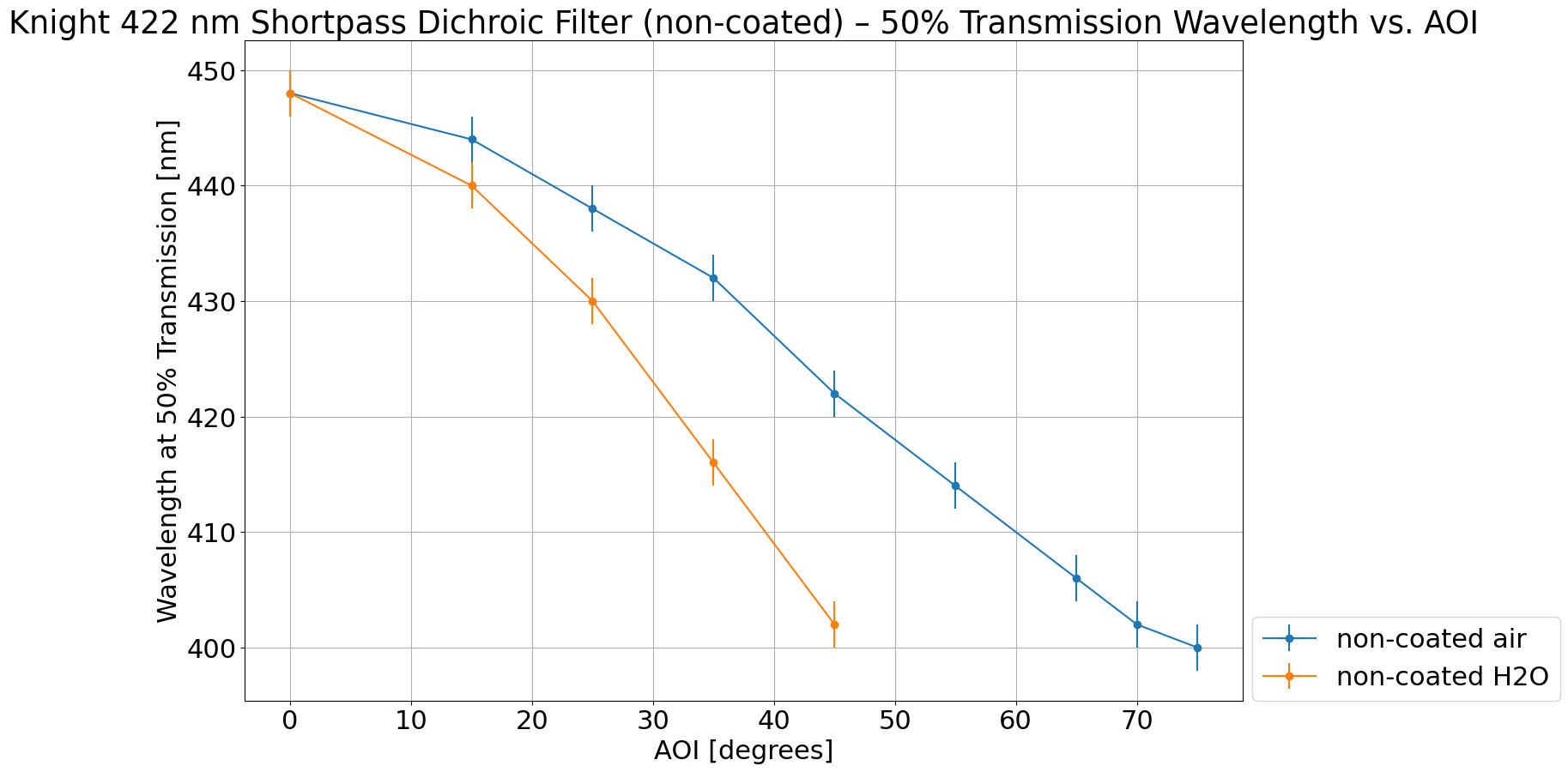}
        \caption{Wavelength at 50$\%$ transmission for the 422~nm shortpass EOS filter (non-coated side) from Knight Optical, measured in air and tap water for AOIs between 0$^\circ$ and 75$^\circ$ using the spectrophotometer. Error bars are 2~nm.}
        \label{fig:Knight_shortpass_EOS_450nm_air_vs_h2o_wave_50_percent_non_coated_error_paper}
    \end{figure}

    \vspace{1cm} 

    \begin{figure}[p!]
        \centering
        \includegraphics[trim=0.2cm 0.2cm 0.2cm 0.2cm, clip=true, width=0.54\textwidth]{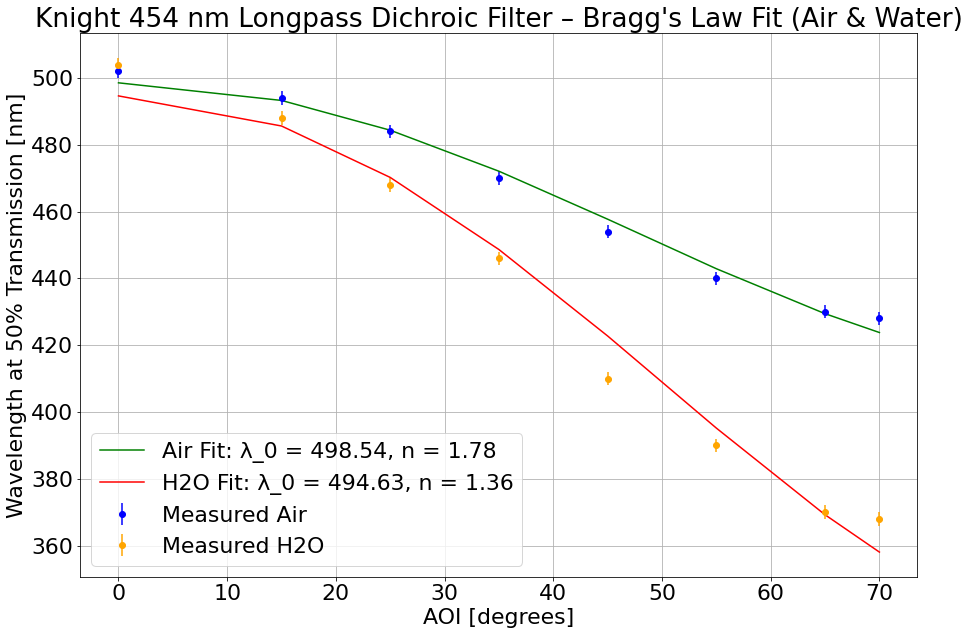}
        \caption{Wavelength at 50$\%$ transmission for the 454~nm longpass filter from Knight Optical, measured in air and tap water for AOIs between 0$^\circ$ and 70$^\circ$ using the spectrophotometer. The data points (error bars: 2~nm) are fitted using the modified Bragg's Law, which accurately models the observed AOI dependence.}
        \label{fig:Knight_500nm_longpass_air_h2o_wave_50_percent_crude_fit_Bragg_bounds_sigma_chi_paper}
    \end{figure}
    
    \vspace{1cm} 
    \newpage
    \FloatBarrier 
}

\afterpage{
    \begin{figure}[p!]
        \centering
        \includegraphics[trim=0.2cm 0.2cm 0.2cm 0.2cm, clip=true, width=0.60\textwidth]{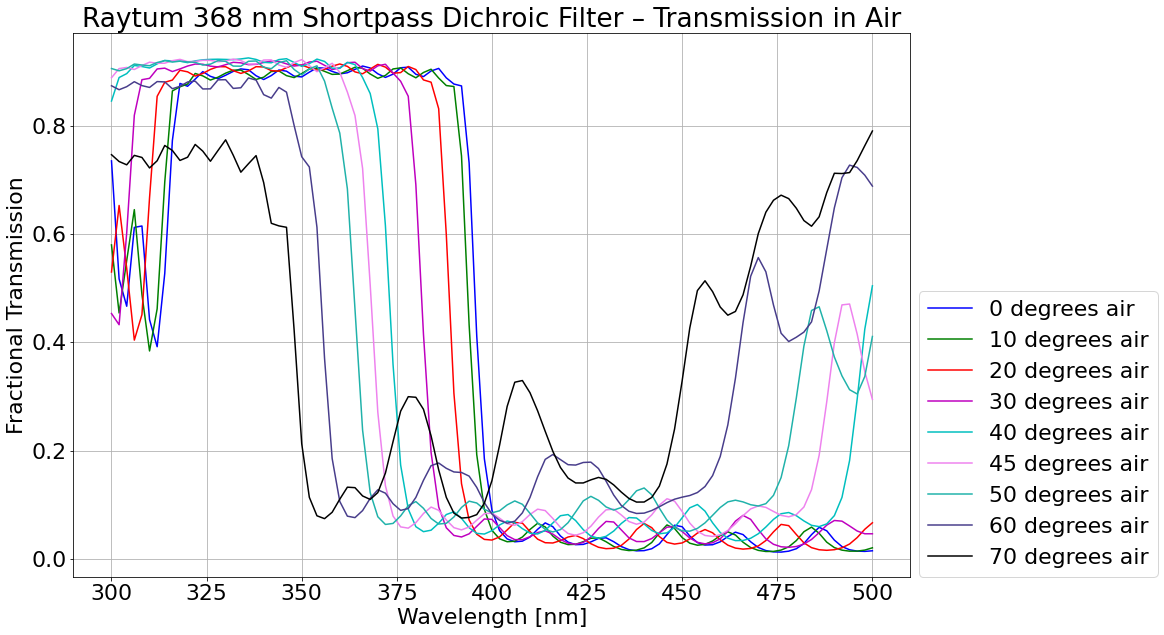}
        \caption{Transmission data (various colors) for the 368~nm shortpass filter from Raytum Photonics, measured in air for AOIs between 0$^\circ$ and 70$^\circ$ using the spectrophotometer.}
        \label{fig:RAYTUM_4_in_SP_air_no_cuvette_coated_2024_07_05_paper}
    \end{figure}

    \vspace{1cm} 

    \begin{figure}[p!]
        \centering
        \includegraphics[trim=0.2cm 0.2cm 0.2cm 0.2cm, clip=true, width=0.60\textwidth]{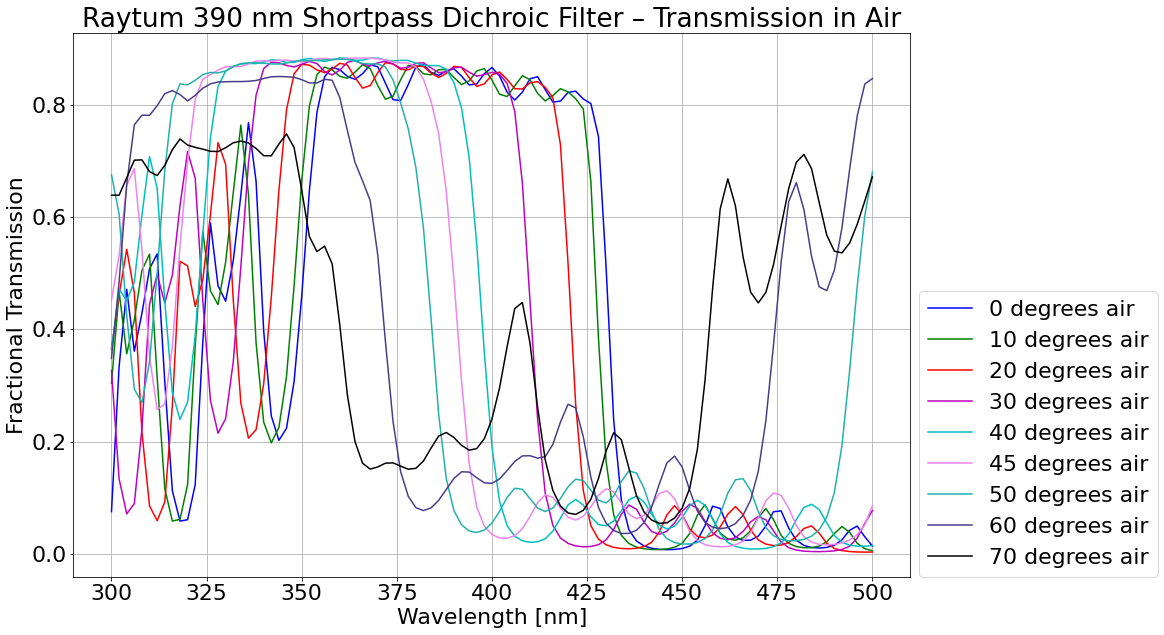}
        \caption{Transmission data (various colors) for the 390~nm shortpass filter from Raytum Photonics, measured in air for AOIs between 0$^\circ$ and 70$^\circ$ using the spectrophotometer.}
        \label{fig:RAYTUM_B15_SP_air_no_cuvette_coated_2024_07_05_paper}
    \end{figure}

    \vspace{1cm} 

    \begin{figure}[p!]
        \centering
        \includegraphics[trim=0.2cm 0.2cm 0.2cm 0.2cm, clip=true, width=0.60\textwidth]{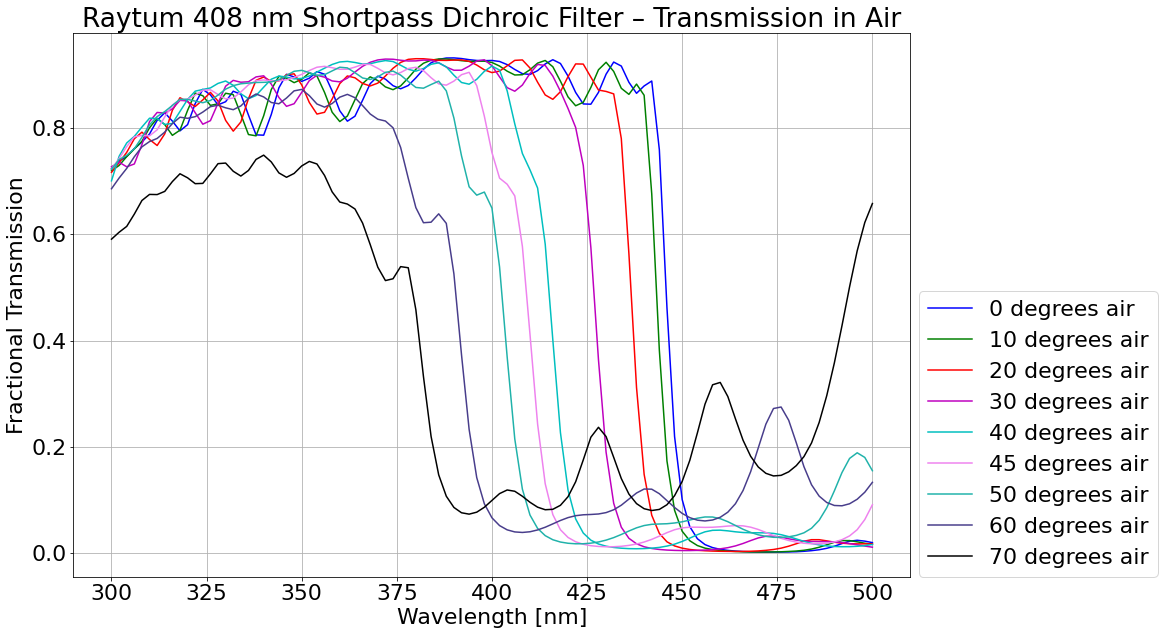}
        \caption{Transmission data (various colors) for the 408~nm shortpass filter from Raytum Photonics, measured in air for AOIs between 0$^\circ$ and 70$^\circ$ using the spectrophotometer.}
        \label{fig:RAYTUM_B23_SP_air_no_cuvette_coated_2024_07_05_paper}
    \end{figure}
    
    \vspace{1cm} 
    \newpage
    \FloatBarrier 
}

\afterpage{
    \begin{figure}[p!]
        \centering
        \includegraphics[trim=0.2cm 0.2cm 0.2cm 0.2cm, clip=true, width=0.54\textwidth]{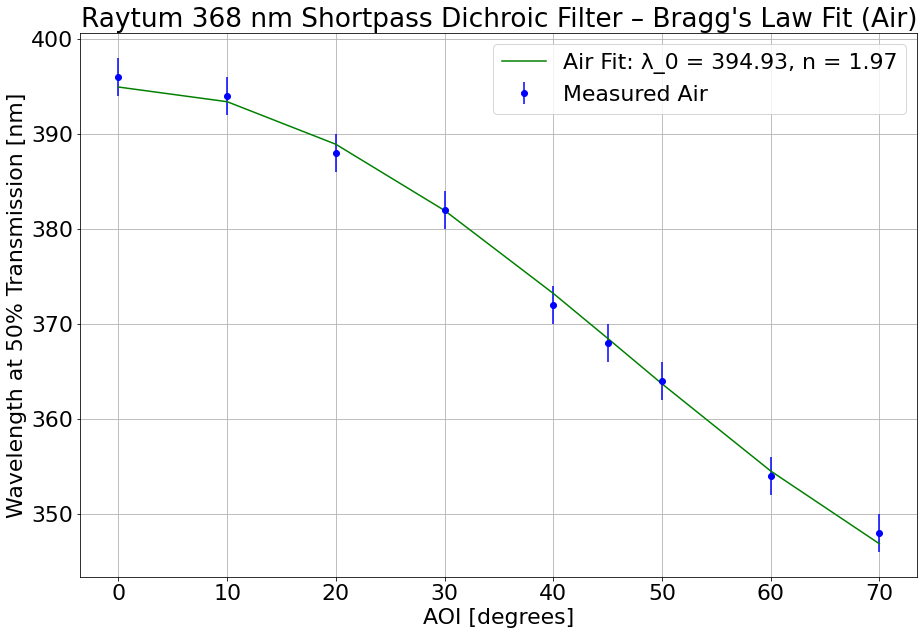}
        \caption{Wavelength at 50$\%$ transmission for the 368~nm shortpass filter from Raytum Photonics, measured in air for AOIs between 0$^\circ$ and 70$^\circ$ using the spectrophotometer. The data points (error bars: 2~nm) are fitted using the modified Bragg's Law, which accurately models the observed AOI dependence.}
        \label{fig:Bragg_4_in_paper}
    \end{figure}

    \vspace{1cm} 

    \begin{figure}[p!]
        \centering
        \includegraphics[trim=0.2cm 0.2cm 0.2cm 0.2cm, clip=true, width=0.54\textwidth]{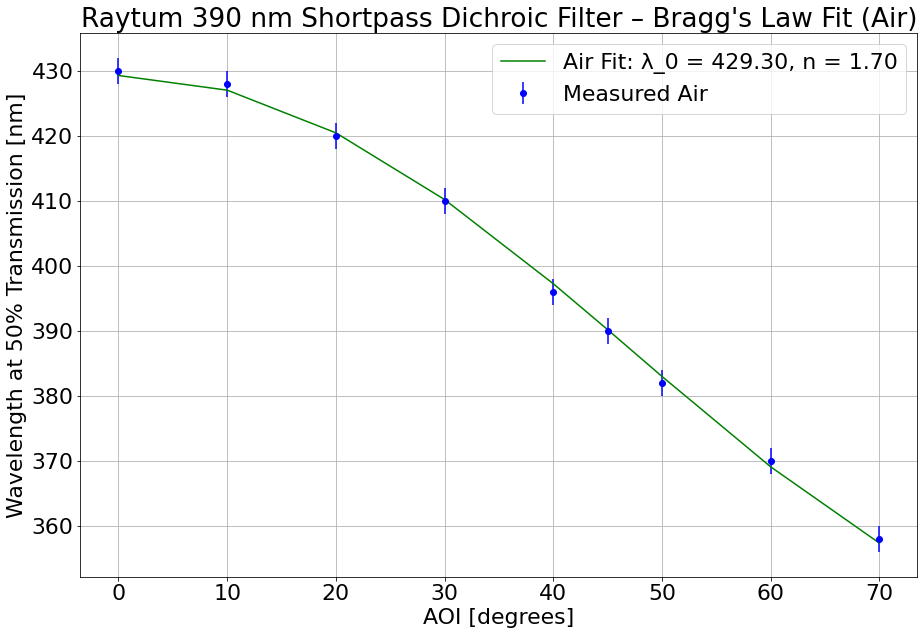}
        \caption{Wavelength at 50$\%$ transmission for the 390~nm shortpass filter from Raytum Photonics, measured in air for AOIs between 0$^\circ$ and 70$^\circ$ using the spectrophotometer. The data points (error bars: 2~nm) are fitted using the modified Bragg's Law, which accurately models the observed AOI dependence.}
        \label{fig:Bragg_B15_paper}
    \end{figure}

    \clearpage


    \begin{figure}[p!]
        \centering
        \includegraphics[trim=0.2cm 0.2cm 0.2cm 0.2cm, clip=true, width=0.54\textwidth]{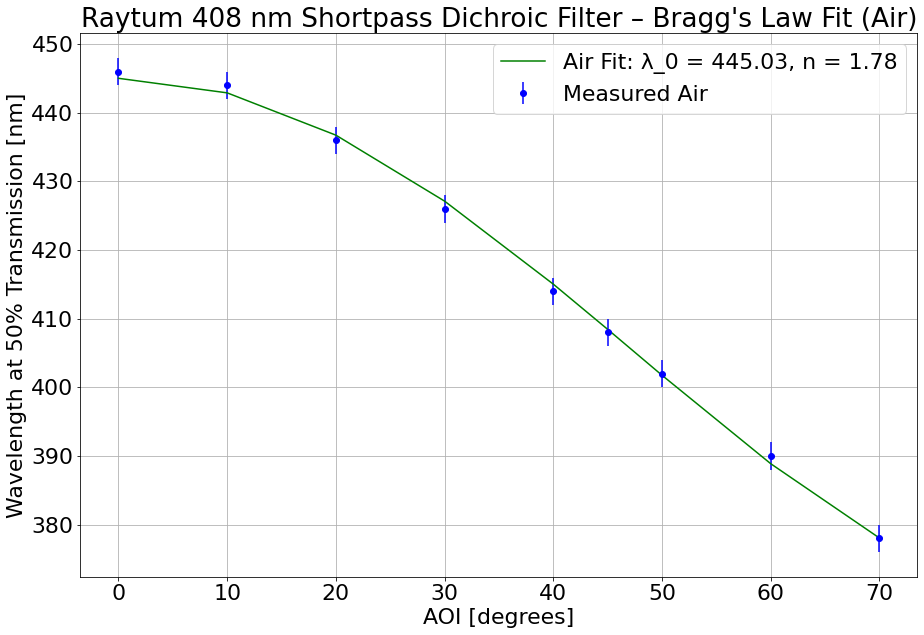}
        \caption{Wavelength at 50$\%$ transmission for the 408~nm shortpass filter from Raytum Photonics, measured in air for AOIs between 0$^\circ$ and 70$^\circ$ using the spectrophotometer. The data points (error bars: 2~nm) are fitted using the modified Bragg's Law, which accurately models the observed AOI dependence.}
        \label{fig:Bragg_B23_paper}
    \end{figure}
    
    \vspace{1cm} 
    \newpage
    \FloatBarrier 
}

\section*{Acknowledgments}

We thank Sam Naugle, Meng Luo, James Shen, and Tanner Kaptanoglu for the useful discussions regarding scintillators and dichroic filter characterization and simulation results. Our thanks to Tony LaTorre and Benjamin Land for developing and updating the LeCrunch and LeCrunch2 software used for the data acquisition. The 3D-printed object is printed with the MakerBot Print and assistance from Michael Reilly at David Rittenhouse Laboratory. This work was supported by the Department of Energy, Office of High Energy Physics, under grant number DE-FG02-88ER40479.

\clearpage

\bibliographystyle{unsrtnat}
\bibliography{bibliography}

\end{document}